\documentclass[fp,twocolumn]{jpsj3}
\usepackage{txfonts}
\usepackage{url}
\usepackage{bm}
\usepackage{color}
\title{Structural Comparison of Error Mitigation Methods for Ising Machines: Penalty-Spin Model versus Stacked Model}

\author{
Tetsuro Abe$^1$,
Kanta Hino$^1$,
and 
Shu Tanaka$^{1,2,3,4,5}\thanks{shu.tanaka@keio.jp}$
}

\inst{
$^1$Graduate School of Science and Technology, Keio University, Kanagawa
223-8522, Japan \\
$^2$Department of Applied Physics and Physico-Informatics, Keio University, Kanagawa 223-8522, Japan \\
$^3$Keio University Sustainable Quantum Artificial Intelligence Center (KSQAIC), Keio University, Tokyo 108-8345,
Japan \\
$^4$Human Biology-Microbiome-Quantum Research Center (WPI-Bio2Q), Keio University, Tokyo 108-8345, Japan\\
$^5$Green Computing System Research Organization, Waseda University, Shinjuku, Tokyo 162-0042, Japan
} %\\

\abst{
Error-mitigation methods for Ising machines are reexamined not merely as  noise-suppression techniques but as a structural design problem of replica-coupled Ising models. 
Using simulated annealing as a hardware-noise-free testbed, we systematically compare the penalty-spin (PS) model, which couples replicas through a centralized auxiliary layer, with the stacked model, which couples adjacent replicas directly. 
Numerical experiments on the quadratic assignment problem reveal that the ferromagnetically coupled stacked model stably maintains constraint satisfaction and improves solution quality over a broad parameter range, exhibiting favorable scalability with both the number of replicas and problem size. 
In contrast, the PS model suffers from cooperation collapse at large parallelism: many-replica averaging in the PS layer washes out sparse solution information, preventing effective inter-replica coordination. These findings demonstrate that the topology of inter-replica couplings decisively influences search robustness, and provide practical guidelines for model selection and parameter tuning in constrained optimization.
}

\begin{document}
\maketitle

\section{Introduction}

Ising machines address combinatorial optimization problems by encoding them into an Ising Hamiltonian and searching for its ground state through physical or physics-inspired dynamics.~\cite{lucas2014ising, mohseni2022ising,tanaka2017quantum,tanahashi2019application}
Using Ising machines, a wide variety of combinatorial optimization problems have been investigated in recent years.~\cite{neukart2017traffic,ohzeki2019control,kitai2020designing,tamura2021performance,inoue2022towards,yarkoni2022quantum,sampei2023quantum,terui2025collaborative,ikeuchi2025evaluating,kikuchi2026high}
Representative annealing-based approaches include quantum annealing (QA), which exploits quantum fluctuations induced by a transverse field, and simulated annealing (SA), which performs a similar optimization using thermal fluctuations.~\cite{kadowaki1998quantum,das2005quantum,das2008colloquium,hauke2020perspectives,chakrabarti2023quantum,kirkpatrick1983optimization,johnson1989optimization,johnson1991optimization}
In the ideal asymptotic limit, convergence to the ground state is guaranteed for QA by the adiabatic theorem and for SA by the Geman--Geman theorem.~\cite{kato1950adiabatic,morita2007convergence,morita2008mathematical,geman1984stochastic}
In practice, however, finite-time annealing inevitably gives rise to nonadiabatic excitations and residual thermal transitions, and the probability of obtaining the ground state can degrade substantially, especially for problems with rugged energy landscapes.

Quantum annealers, developed as QA-based Ising machines, are analog devices and thus suffer from hardware-implementation errors beyond the intrinsic stochasticity of finite-time annealing.
Because couplings and local fields are realized as continuous physical parameters, unavoidable control errors perturb the intended Hamiltonian.
In random spin systems, even small perturbations of couplings can reshuffle low-energy configurations, a phenomenon known as $J$-chaos,~\cite{Nifle1992new, Nifle1998chaos, Katzgraber2007temperature, Albash2019analog, pearson2019analog} potentially altering the ground-state configuration itself.
Consequently, even when the annealing dynamics itself is ideal, the device may return an optimum of a perturbed (``wrong'') Hamiltonian, severely deteriorating the scaling of search performance.
This ``wrong-Hamiltonian'' issue and the use of stabilizer/repetition-type encodings to suppress the impact of random control errors have been analyzed in the context of adiabatic quantum optimization.~\cite{young2013adiabatic}

To mitigate such effects, error-mitigation frameworks that embed redundancy directly into the Hamiltonian have been proposed, notably quantum annealing correction (QAC).~\cite{pudenz2014error,pudenz2015quantum,vinci2015quantum,vinci2016nested,matsuura2016mean,matsuura2017quantum,vinci2018scalable,matsuura2019nested,pearson2019analog,hattori2025frustration}
Rather than post-processing readout samples,~\cite{kanamaru2021solving} QAC introduces redundancy and additional couplings to suppress the creation and propagation of errors during the annealing process.
A representative realization is the penalty-spin (PS) model,~\cite{pudenz2014error} where multiple replicas of logical variables are indirectly coupled through auxiliary penalty spins, such that mismatches among replicas are penalized energetically.
This centralized coupling creates an energy barrier against errors and experimental studies have reported improvements in success probabilities.~\cite{pudenz2014error,pudenz2015quantum,vinci2015quantum,vinci2016nested,vinci2018scalable,pearson2019analog}

An alternative family of designs couples replicas directly without auxiliary spins, i.e., a repetition/stacked-type construction in which corresponding variables across replicas are linked by inter-replica couplings.~\cite{young2013adiabatic,bennett2023using}
In our stacked model, we focus on a decentralized topology with interactions only between neighboring replicas, which eliminates the need for a central auxiliary layer.
Moreover, by tuning the sign and magnitude of inter-replica couplings, one can promote synchronization while retaining partial independence among replicas.
Although both PS-type and stacked-type constructions were originally motivated by error mitigation, their fundamentally different interaction topologies alter the effective energy landscape and the search dynamics in nontrivial ways.

A key difficulty in much of the existing literature is disentangling genuine noise suppression from algorithmic or structural effects induced by the additional couplings. 
Bennett et al.~\cite{bennett2023using} addressed this issue by analyzing ground-state properties and quantum-walk dynamics under a precision-limited error model, demonstrating that weak antiferromagnetic links with frustration can improve error tolerance for Sherrington--Kirkpatrick spin glasses.
However, the behavior under finite-time annealing dynamics, which is more directly relevant to practical quantum annealing and classical simulated annealing implementations, has not been systematically investigated.
Strong inter-replica couplings can also over-constrain replicas and induce collective freezing, where multiple replicas become trapped together in suboptimal local minima.~\cite{tanaka2009mechanism,kikuchi2023dynamical,abe2025error}
These dynamical risks are expected to depend on both the nature and the scale of the problem, and they can be particularly pronounced for constrained optimization problems with highly sparse feasible solutions.

In this work, we employ SA as a hardware-noise-free testbed to isolate and systematically compare the structural effects of different replica-coupling topologies on search dynamics.
We reframe error mitigation not merely as a remedy for hardware noise but as a structural design problem of replica-coupled Ising models.
Here, the term ``error mitigation'' is used in a broader sense, referring to algorithmic and structural mechanisms that improve the robustness of the search process, not solely to suppression of hardware noise.
By providing a noise-free baseline for finite-time dynamics, our analysis clarifies topology-induced cooperation and failure modes that are otherwise entangled with hardware errors.

As a benchmark, we focus on the quadratic assignment problem (QAP),~\cite{koopmans1957assignment} a constrained combinatorial optimization problem with strongly correlated constraints.
In typical QAP formulations, feasible solutions are highly sparse, which can make naive replica-consistency measures (e.g., simple spin correlations) misleading.
Through numerical analyses on QAP instances, we (i) map stable operating regions with respect to the constraint-penalty coefficient and the inter-replica coupling, (ii) elucidate mechanisms of replica coordination and failure under sparse constraints, and (iii) compare scalability with respect to problem size and the number of replicas.
Through these analyses, we evaluate error-mitigation methods from the perspective of structural design and clarify the applicability and limitations of the PS and stacked models.

The remainder of this paper is organized as follows.
In Sect.~\ref{Sec:modelsandproblem}, we define the replica-coupled Ising models, the independent-replica model (C model), the PS model, and the stacked model, and formulate the QAP as an Ising Hamiltonian.
Section~\ref{Sec:simulationmethods} describes the simulation setup, including the SA algorithm and the temperature schedule.
In Sect.~\ref{Sec:results}, we present numerical results on QAP instances, examining the dependence on the penalty coefficient, the inter-replica coupling strength, the number of replicas, and the problem size.
Section~\ref{Sec:mechanism} analyzes the mechanisms underlying the observed performance differences, focusing on inter-replica correlations and bit-configuration statistics.
Based on these findings, Sect.~\ref{Sec:practicalguidelines} provides practical guidelines for model selection and parameter tuning.
Finally, Sect.~\ref{Sec:conclusion} concludes this paper with a summary and an outlook for future work.

\section{Models and Problem}
\label{Sec:modelsandproblem}

In this section, we define the replica-coupled Ising models evaluated in this work and describe the target optimization problem. 
We first review the structures of the penalty-spin (PS) model and the stacked model, which were originally proposed for error mitigation, and clarify the framework for comparison.
We then present an Ising formulation of the quadratic assignment problem (QAP).

\subsection{Replica-coupled Ising models}
\label{subsec:replica-coupledIsingmodels}

The basic idea behind error-mitigation methods is to create multiple replicas of the original Ising model and introduce inter-replica couplings, thereby enhancing the robustness of the search.
In this study, robustness does not refer to resilience against thermal noise or control errors, but rather to the stability of constraint satisfaction and solution quality under variations in the structure and parameters of the inter-replica interactions.

We prepare $P$ layers in total. 
In the C model and the stacked model, each layer $p=1,\dots,P$ is a problem replica governed by the same QAP Hamiltonian $H_0^{(p)}(\bm{s}^{(p)})$. 
In contrast, in the PS model, only $P-1$ layers ($p=1,\dots,P-1$) are problem replicas, while the remaining layer ($p = P$) serves as an auxiliary penalty-spin (PS) layer and does not contain $H_0$.

\begin{figure}[t]
    \centering
    \includegraphics[width=\linewidth]{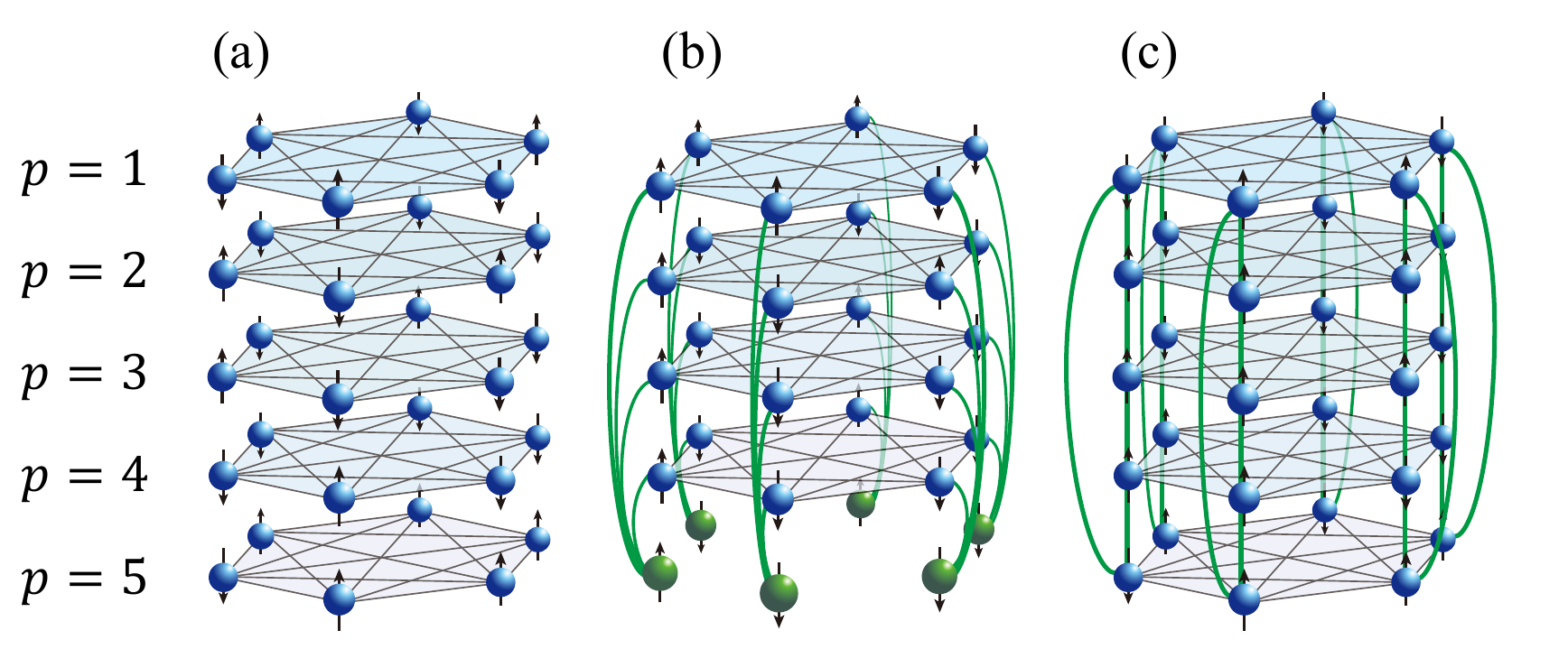}
    \caption{(Color online) Schematic illustrations of replica-coupled models with $P=5$ replicas:
    (a) independent-replica model (C model),
    (b) penalty-spin (PS) model, and
    (c) stacked model.}
    \label{Fig:errormitigationmodels}
\end{figure}

Here, $\bm{s}^{(p)}=(s^{(p)}_{1},\ldots,s^{(p)}_{N})$ denotes the spin variables in the $p$-th replica, and $N$ is the number of spins per replica.
Below, we define the three models compared in this work (Fig.~\ref{Fig:errormitigationmodels}).

The independent-replica model (C model), illustrated in Fig.~\ref{Fig:errormitigationmodels}(a), consists of $P$ noninteracting replicas.
The total Hamiltonian is
\begin{align}
    H_{\rm C}(P) = \sum_{p=1}^{P} H_0^{(p)}\!\left(\bm{s}^{(p)}\right),
\end{align}
with no couplings between replicas.
This model corresponds to simple parallel execution and serves as a baseline for evaluating the performance gains from inter-replica couplings.

In the penalty-spin (PS) model, shown in Fig.~\ref{Fig:errormitigationmodels}(b), one of the $P$ layers serves as an auxiliary layer.
Although this layer does not include the problem Hamiltonian $H_0$, it still represents a configuration in the original spin space and can therefore be evaluated a posteriori using $H_0$.
The Hamiltonian is
\begin{align}
    H_{\rm PS}(P)
    = H_{\rm C}(P-1)
    + J_{\rm P} \sum_{p=1}^{P-1}\sum_{i=1}^{N} s_{i}^{(p)} s_{i}^{(P)},
\end{align}
where $J_{\rm P}$ is the inter-replica coupling strength.
As indicated by the green links in Fig.~\ref{Fig:errormitigationmodels}(b), each problem replica ($p=1,\ldots,P-1$) interacts only with the PS layer ($p=P$). 
There are no direct couplings among problem replicas, and correlations are mediated through the PS layer.

The stacked model, shown in Fig.~\ref{Fig:errormitigationmodels}(c), couples replicas directly without auxiliary spins.
It adopts a ring topology in replica space, corresponding spins in adjacent replicas interact with periodic boundary conditions, $s_{i}^{(P+1)}=s_{i}^{(1)}$ for all $i$.
The Hamiltonian is
\begin{align}
    H_{\rm Stacked}(P)
    = H_{\rm C}(P)
    + J_{\rm P} \sum_{p=1}^{P}\sum_{i=1}^{N} s_{i}^{(p)} s_{i}^{(p+1)}.
\end{align}

In this study, comparisons between different models are made at a fixed total number of layers $P$, which is used as a proxy for computational resources, reflecting the total number of spins and couplings required in a physical implementation.
Although the number of problem replicas differs between the PS model and the other models, the total number of layers (and hence the total number of spins) is kept identical.
In the PS model, one layer is explicitly allocated as an auxiliary penalty-spin layer rather than a problem replica. 
This overhead is an intrinsic structural cost of the model design rather than an artifact of the comparison.
Accordingly, differences in performance at fixed $P$ reflect genuine consequences of the coupling topology, not merely the number of problem replicas.

In analog Ising machines, increasing $P$ primarily increases the required number of physical spins, whereas the annealing time per run is essentially unchanged because all spins evolve in parallel.
In digital implementations, the wall-clock overhead can also be substantially smaller than a naive $P$-fold increase when parallel updates across layers are available, although the exact scaling depends on the specific hardware and software architecture.

We note that the use of multiple replicas in this work is conceptually distinct from replica exchange or parallel tempering Monte Carlo methods.~\cite{hukushima1996exchange,hansmann1997parallel,earl2005parallel}
In replica exchange approaches, replicas evolve independently at different temperatures and occasionally exchange configurations to enhance equilibration.
By contrast, the replica-coupled models studied here involve explicit inter-replica interactions at a single temperature, through which cooperation or competition among replicas is continuously mediated during the dynamics.
As a result, the mechanisms and objectives of replica coupling considered here differ fundamentally from temperature-based replica exchange schemes.

\subsection{Decoding}
\label{subsec:decoding}

\begin{figure}[t]
    \centering
    \includegraphics[width=\linewidth]{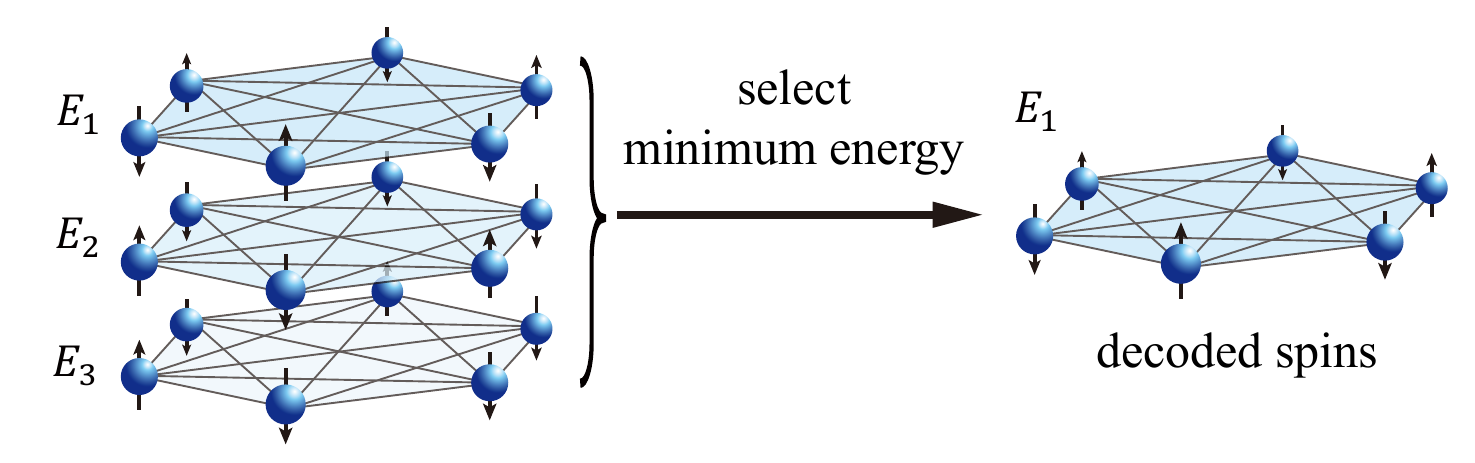}
    \caption{Schematic of minimum-energy decoding: the replica with the lowest energy under $H_0$ is selected as the output.}
    \label{Fig:minimum_energy_decode}
\end{figure}

After running SA on a replica-coupled model, we obtain $P$ spin configurations $\bm{s}^{(p)}$ corresponding to the $P$ layers ($p=1,\dots,P$).
To extract a single solution from the $P$ layers, we employ \emph{minimum-energy decoding} with respect to the original problem Hamiltonian $H_0$ (see Fig.~\ref{Fig:minimum_energy_decode}).
This decoding treats all layers, including the PS layer, as candidate solutions by evaluating their configurations under $H_0$, regardless of whether $H_0$ explicitly appears in the dynamics.
The procedure is as follows:
\begin{enumerate}
    \item Compute the energy of each replica under the original Hamiltonian:
    \begin{align}
        E_p = H_0\!\left(\bm{s}^{(p)}\right).
    \end{align}
    \item Select the replica $p^\ast$ with the lowest energy.
    \item Output $\bm{s}^{(p^\ast)}$ as the final solution.
\end{enumerate}
In summary,
\begin{align}
    \bm{s}_{\mathrm{final}} = \bm{s}^{(p^\ast)}, \quad
    p^\ast = \underset{p \in \{1,\dots,P\}}{\operatorname*{arg\,min}}\;
    H_0\!\left(\bm{s}^{(p)}\right).
\end{align}
We emphasize that the purpose of employing minimum-energy decoding is not to advocate a specific decoding strategy, but to provide a decoder-agnostic and uniform evaluation rule applicable to all replica-coupled models.
By evaluating all layers using the same original problem Hamiltonian $H_0$, this decoding allows a fair comparison of solution quality independent of the details of the underlying solver dynamics.
In the PS model, although the penalty-spin layer is not directly optimized with respect to $H_0$ during annealing, it still represents a configuration in the original variable space and can therefore be evaluated a posteriori under $H_0$ on the same footing as the problem replicas.

\subsection{Quadratic assignment problem}
\label{subsec:QAP}

The quadratic assignment problem (QAP) is a combinatorial optimization problem in which $L$ facilities are assigned one-to-one to $L$ locations so as to minimize a total cost that depends on inter-facility flows and inter-location distances.~\cite{koopmans1957assignment}
Let $f_{i,j}$ denote the flow from facility $i$ to facility $j$, and let $d_{k,l}$ denote the distance between locations $k$ and $l$.
For an assignment represented by a permutation $\pi$ (facility $i$ is placed at location $\pi(i)$), the objective function is
\begin{align}
\label{Eq:qap_objective_originalform}
    C(\pi) = \sum_{i=1}^{L}\sum_{j=1}^{L} f_{i,j}\, d_{\pi(i),\pi(j)}.
\end{align}
The QAP is a well-known NP-hard problem. 
The number of feasible assignments grows as ${\cal O}(L!)$, making exhaustive search quickly intractable.~\cite{sahni1976p}

In this study, we first express the QAP in quadratic unconstrained binary optimization (QUBO) form and then convert it to an Ising Hamiltonian via a standard transformation.
We introduce binary variables $x_{i,k}\in\{0,1\}$, where $x_{i,k}=1$ if facility $i$ is assigned to location $k$ and $x_{i,k}=0$ otherwise.
Using these variables, the objective function~\eqref{Eq:qap_objective_originalform} becomes
\begin{align}
\label{Eq:qap_objective_quboform}
    H_{\rm obj}(\bm{x})
    = \sum_{i=1}^{L}\sum_{j=1}^{L}\sum_{k=1}^{L}\sum_{l=1}^{L}
    f_{i,j}\, d_{k,l}\, x_{i,k}\, x_{j,l}.
\end{align}

The one-to-one assignment imposes two sets of constraints.
First, each facility must be assigned to exactly one location:
\begin{align}
\label{Eq:constraint_facility}
    \sum_{k=1}^{L} x_{i,k} = 1, \quad \forall\, i.
\end{align}
Second, each location must accommodate exactly one facility:
\begin{align}
\label{Eq:constraint_place}
    \sum_{i=1}^{L} x_{i,k} = 1, \quad \forall\, k.
\end{align}
We encode these constraints as a quadratic penalty term that vanishes if and only if all constraints are satisfied:
\begin{align}
\label{Eq:qap_constraint_quboform}
    H_{\rm const}(\bm{x})
    = \sum_{i=1}^{L}\left(\sum_{k=1}^{L} x_{i,k}-1\right)^{\!2}
    + \sum_{k=1}^{L}\left(\sum_{i=1}^{L} x_{i,k}-1\right)^{\!2}.
\end{align}
A solution satisfying Eqs.~\eqref{Eq:constraint_facility} and \eqref{Eq:constraint_place} is called a \emph{feasible solution}.
The QUBO Hamiltonian is then
\begin{align}
\label{Eq:QAP_quboform}
    H_{\rm QAP}(\bm{x}) = H_{\rm obj}(\bm{x}) + \mu\, H_{\rm const}(\bm{x}),
\end{align}
where $\mu>0$ is the penalty coefficient.

A characteristic feature of the QAP formulation is its sparsity: in any feasible solution, exactly $L$ of the $N = L^{2}$ variables take the value~1, while all others are~0.
This extreme sparsity plays an important role in evaluating inter-replica coordination, as discussed in Sect.~\ref{Sec:mechanism}.
We emphasize that QAP is not chosen merely as a classical benchmark, but as a representative example of constrained optimization problems with sparse feasible structures.
In many real-world optimization tasks, constraints such as assignment, matching, scheduling, and capacity constraints effectively restrict solutions to a small subset of variables being active, often resulting in one-hot or near one-hot representations.~\cite{lucas2014ising,tamura2021performance,kikuchi2024performance,ide2025extending}
Therefore, understanding how replica-coupling mechanisms behave under sparse feasible structures is of broad relevance beyond QAP itself.

In this study, we normalize the flow and distance matrices by their respective maxima,
$f_{i,j}\leftarrow f_{i,j}/\max_{i',j'} f_{i',j'}$ and $d_{k,l}\leftarrow d_{k,l}/\max_{k',l'} d_{k',l'}$,
so that the coefficient scales are uniform across problem instances.

Because the replica-coupling terms are most naturally expressed in terms of Ising spins, we convert the QUBO to an Ising formulation by mapping $x_{i,k}\in\{0,1\}$ to $s_{i,k}\in\{+1,-1\}$ via
\begin{align}
\label{Eq:binary_to_spin}
    x_{i,k} = \frac{s_{i,k}+1}{2}.
\end{align}
The resulting Ising Hamiltonian (up to an additive constant), denoted $H_0(\bm{s})$, serves as the base Hamiltonian for all replica-coupled models defined above.

\subsection{Evaluation metrics}
\label{subsec:metrics}

We use the following metrics to evaluate the performance of each model quantitatively.

\paragraph{Approximation ratio.}
Here, $C_{\mathrm{sol}}$ denotes the objective value of a decoded solution evaluated using the normalized objective function defined in Eq.~\eqref{Eq:qap_objective_originalform} (equivalently Eq.~\eqref{Eq:qap_objective_quboform}), i.e., excluding the constraint-penalty term, and $C_{\mathrm{opt}}$ is the corresponding optimal objective value.
The value $C_{\mathrm{opt}}$ is obtained by evaluating a known optimal assignment reported in QAPLIB under the same normalization and variable mapping as used throughout this study.
For a feasible solution with objective value $C_{\mathrm{sol}}$, the approximation ratio to the known optimum $C_{\mathrm{opt}}$ is defined as
\begin{equation}
    R = \frac{C_{\mathrm{sol}}}{C_{\mathrm{opt}}}.
\end{equation}
The value $R$ is computed over feasible samples only.
If no feasible solutions are obtained, $R$ is undefined and not plotted.
Since the QAP is a minimization problem, $R \ge 1$ holds, and values closer to $1$ indicate better solution quality.

\paragraph{Feasibility rate.}
The feasibility rate $P_{\mathrm{Feasible}}$ is the fraction of samples satisfying all QAP constraints:
\begin{equation}
    P_{\mathrm{Feasible}} = \frac{N_{\mathrm{Feasible}}}{N_{\mathrm{All}}},
\end{equation}
where $N_{\mathrm{All}}$ is the total number of samples and $N_{\mathrm{Feasible}}$ is the number of feasible samples.

\paragraph{Mean bit value.}
The mean bit value measures the fraction of variables taking the value~$1$.
For a set of replicas $\mathcal{P}$, it is defined as
\begin{equation}
    \langle x \rangle_{\mathcal{P}}
    = \frac{1}{|\mathcal{P}|\, N_{\mathrm{All}}\, N}
      \sum_{p \in \mathcal{P}}
      \sum_{n=1}^{N_{\mathrm{All}}}
      \sum_{i,k}
      x_{i,k}^{(n,p)},
\end{equation}
where $|\mathcal{P}|$ is the number of replicas in $\mathcal{P}$, $x_{i,k}^{(n,p)} \in \{0,1\}$ is the value of variable $(i,k)$ in replica $p$ for the $n$-th sample, and $N = L^{2}$ is the number of variables per replica.
We set $\mathcal{P} = \{1,\dots,P\}$ when averaging over all replicas and $\mathcal{P} = \{P\}$ when focusing on the PS layer alone.

\section{Simulation Methods}
\label{Sec:simulationmethods}

To isolate the effect of inter-replica coupling topology on search dynamics, we employ simulated annealing (SA) as a hardware-noise-free testbed.
We use \texttt{SASampler} from the open-source Python library OpenJij (version 0.10.6),~\cite{openjij} which implements a Markov-chain Monte Carlo (MCMC) algorithm based on the Metropolis rule.~\cite{metropolis1953equation,hastings1970monte}
The same update rule is applied to all models considered in this study.

For the temperature schedule, we adopt OpenJij's default geometric progression in inverse temperature.
The aim is to allow broad exploration at high temperature and then spend sufficient time in the low-temperature regime to promote relaxation toward near-equilibrium states.
Let $N_{\mathrm{Steps}}$ denote the total number of MCMC steps.
The inverse temperature at step $t$ ($1 \le t \le N_{\mathrm{Steps}}$) is
\begin{align}
    \beta_t
    = \beta_{\mathrm{init}}
      \left( \frac{\beta_{\mathrm{final}}}{\beta_{\mathrm{init}}} \right)^{t/N_{\mathrm{Steps}}}.
\end{align}

The initial and final inverse temperatures, $\beta_{\mathrm{init}}$ and $\beta_{\mathrm{final}}$, are determined adaptively for each problem instance following OpenJij's standard procedure.
Before running SA, we generate two independent random spin configurations and evaluate the energy change $\Delta E$ for every single-spin flip, yielding $2N$ samples.
From the positive values of $\Delta E$, we take the median as $\Delta E_{\mathrm{typ}}$ and the minimum as $\Delta E_{\mathrm{min}}$.
Here, $\Delta E_{\mathrm{typ}}$ serves as a proxy for a typical energy barrier, while $\Delta E_{\mathrm{min}}$ represents the smallest positive energy increment.

The initial inverse temperature $\beta_{\mathrm{init}}$ is chosen to ensure sufficient thermal fluctuations for global exploration, targeting an average acceptance probability of $P_{\mathrm{init}} \approx 0.9$.
A reference value $\beta_{\mathrm{base}}$ is obtained by binary search so as to satisfy
\begin{align}
    2 \beta_{\mathrm{base}} P_{\mathrm{init}}
    \approx
    e^{-\beta_{\mathrm{base}}\Delta E_{\mathrm{min}}}
    -
    e^{-\beta_{\mathrm{base}}\Delta E_{\mathrm{typ}}}.
\end{align}
A correction accounting for the number of spins $N$ and the number of steps $N_{\mathrm{Steps}}$ is then applied:
\begin{align}
    \beta_{\mathrm{init}}
    =
    0.8\,\beta_{\mathrm{base}}
    +
    0.7\,\beta_{\mathrm{base}}
    \tanh\!\left(\frac{N_{\mathrm{Steps}}}{N}\right).
\end{align}

The final inverse temperature $\beta_{\mathrm{final}}$ is set so that acceptance probabilities become sufficiently small at the end of the annealing, using $P_{\mathrm{final}} \le 10^{-3}$ as a guideline.
Let $J_{\mathrm{min}}$ denote the smallest absolute value among the interaction coefficients in the problem Hamiltonian.
We compute
\begin{align}
    \beta_{\Delta E} = -\frac{\ln P_{\mathrm{final}}}{\Delta E_{\mathrm{min}}},
    \qquad
    \beta_{J} = -\frac{\ln P_{\mathrm{final}}}{J_{\mathrm{min}}},
\end{align}
and set
\begin{align}
    \beta_{\mathrm{final}}
    =
    \beta_{\mathrm{small}}
    +
    \left(\beta_{\mathrm{large}}-\beta_{\mathrm{small}}\right)
    \tanh\!\left(\frac{N_{\mathrm{Steps}}}{2N}\right),
\end{align}
where $\beta_{\mathrm{small}}=\min(\beta_{\Delta E},\beta_{J})$ and $\beta_{\mathrm{large}}=\max(\beta_{\Delta E},\beta_{J})$.

For each parameter setting, we perform $1000$ independent SA runs starting from different random seeds and report ensemble averages over these trials.
We emphasize that our goal is not to optimize SA performance itself, but rather to compare the effects of different inter-replica coupling topologies under identical search conditions.
Accordingly, the same temperature schedule and sampling parameters are used for all models.
\section{Results: Performance, Robustness, and Size Scaling on QAP}
\label{Sec:results}

In this section, we report SA-based numerical results for the three replica-coupled models defined in Sect.~\ref{subsec:replica-coupledIsingmodels}: the independent-replica model (C model), the PS model, and the stacked model.
Throughout the numerical comparisons below, the parameter $P$ denotes the total number of layers.
In the PS model, one layer is reserved for the auxiliary penalty-spin layer, whereas in the C and stacked models all layers correspond to problem replicas.
We note that increasing $P$ in the PS model does not simply increase the number of independent problem replicas but also strengthens the averaging effect mediated by the PS layer.
The target problem is the QAP, introduced in Sect.~\ref{subsec:QAP}.
We first analyze a small instance (\texttt{tai12a} from QAPLIB,~\cite{QAPLIBWebsite}) examining the dependence on the penalty coefficient $\mu$ and the inter-replica coupling $J_{\mathrm{P}}$, as well as scalability with respect to $P$ and $N_{\mathrm{Steps}}$.
We then verify that the observed trends persist for larger instances (\texttt{tai15a}, \texttt{tai17a}, and \texttt{tai20a}) through size-scaling analyses.
The main focus is to clarify, in the absence of hardware noise, how the inter-replica coupling topology affects constraint satisfaction and solution quality.

\subsection{Optimization performance on tai12a: dependence on $\mu$}
\label{subsec:mu_dependence}

In this subsection, we evaluate the optimization performance of each model on the 12-location QAPLIB instance \texttt{tai12a}.
Because the QAP Hamiltonian consists of an objective term and a constraint-penalty term, the penalty coefficient $\mu$ critically affects both constraint satisfaction and solution quality.
If $\mu$ is too small, constraint-violating configurations become energetically favorable, making feasible solutions difficult to obtain.
Conversely, if $\mu$ is too large, the penalty term dominates the energy landscape: constraints are satisfied, but optimization of the objective is hindered, degrading the approximation ratio.
Furthermore, the inter-replica coupling $J_{\mathrm{P}}$ in the PS and stacked models directly influences the search dynamics, so the optimal combination of $(\mu, J_{\mathrm{P}})$ is nontrivial.

\begin{figure}[t]
    \centering
    \includegraphics[width=\linewidth]{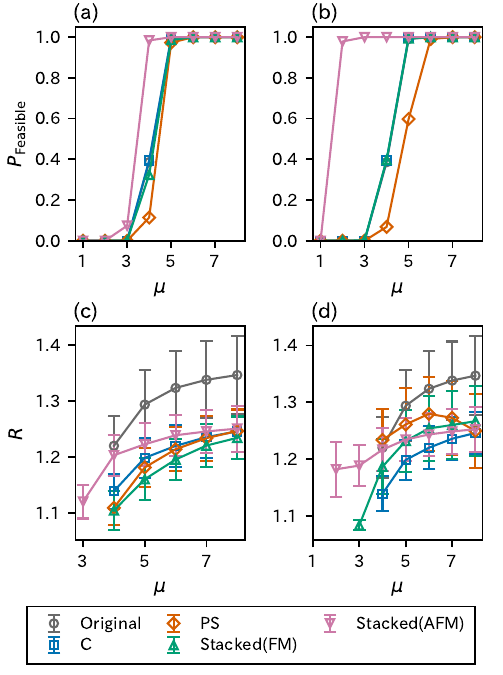}
    \caption{Dependence of $P_{\mathrm{Feasible}}$ and $R$ on $\mu$ for $P=10$.
    (a)~$P_{\mathrm{Feasible}}$ at $|J_{\mathrm{P}}|=0.6$.
    (b)~$P_{\mathrm{Feasible}}$ at $|J_{\mathrm{P}}|=3$.
    (c)~$R$ at $|J_{\mathrm{P}}|=0.6$.
    (d)~$R$ at $|J_{\mathrm{P}}|=3$.
    Error bars indicate standard deviations.
    Lines connecting data points are guides to the eye.}
    \label{Fig:feasible_rate_vs_mu}
\end{figure}

We begin with a preliminary analysis to identify the $\mu$ range required for stable constraint satisfaction.
Two coupling strengths, $|J_{\mathrm{P}}|=0.6$ and $|J_{\mathrm{P}}|=3$, are considered, and we compare the C model, the PS model, and the stacked model with ferromagnetic (FM, $J_{\mathrm{P}}<0$) and antiferromagnetic (AFM, $J_{\mathrm{P}}>0$) couplings.
Results for the original single-replica setting are also shown as a baseline.
Figure~\ref{Fig:feasible_rate_vs_mu} shows the $\mu$ dependence of the feasibility rate $P_{\mathrm{Feasible}}$ and the approximation ratio $R$ for $P=10$: panels~(a) and (c) correspond to $|J_{\mathrm{P}}|=0.6$, and panels~(b) and (d) to $|J_{\mathrm{P}}|=3$.

As shown in Fig.~\ref{Fig:feasible_rate_vs_mu}(a), for $\mu \le 3$ all models except the AFM-stacked model exhibit markedly low $P_{\mathrm{Feasible}}$.
This occurs because the penalty term is too weak, making constraint-violating configurations energetically favorable.
As $\mu$ increases, $P_{\mathrm{Feasible}}$ rises; for $\mu \ge 5$, all models achieve $P_{\mathrm{Feasible}} \approx 1.0$, indicating stable recovery of feasible solutions.
The AFM-stacked model attains the highest $P_{\mathrm{Feasible}}$ at $\mu = 3$, and $4$, suggesting that it satisfies constraints more easily over a broader $\mu$ range.

Under stronger coupling [$|J_{\mathrm{P}}|=3$; Fig.~\ref{Fig:feasible_rate_vs_mu}(b)], the original, C, and FM-stacked models again reach $P_{\mathrm{Feasible}} \approx 1.0$ for $\mu \ge 5$, as in the $|J_{\mathrm{P}}|=0.6$ case.
In contrast, the PS model shows lower $P_{\mathrm{Feasible}}$ than at $|J_{\mathrm{P}}|=0.6$ for the same $\mu$, indicating that its stable operating range depends sensitively on $J_{\mathrm{P}}$.
The AFM-stacked model exhibits a similar trend while maintaining $P_{\mathrm{Feasible}} \approx 1.0$ over a broader $\mu$ range.

Figures~\ref{Fig:feasible_rate_vs_mu}(c) and \ref{Fig:feasible_rate_vs_mu}(d) show the $\mu$ dependence of the approximation ratio $R$.
In Fig.~\ref{Fig:feasible_rate_vs_mu}(c), the original setting exhibits pronounced degradation of $R$ with increasing $\mu$, reflecting the typical trade-off between constraint enforcement and objective optimization.
The C, PS, and stacked models show milder degradation, suggesting that parallelization (introducing replicas) improves robustness of solution quality against $\mu$ variations.
Among these, the FM-stacked model consistently maintains the smallest $R$ over a wide $\mu$ range, followed by the PS and C models.
The AFM-stacked model achieves high $P_{\mathrm{Feasible}}$ but exhibits larger $R$ than the other models, indicating a different type of trade-off between constraint satisfaction and solution quality.

In Fig.~\ref{Fig:feasible_rate_vs_mu}(d), both the PS and FM-stacked models show degraded $R$ over a wide range of $\mu$ values, indicating that, for this parameter setting ($P=10$), $|J_{\mathrm{P}}|=3$ leads to overly strong inter-replica constraints.
The replicas become overly constrained toward identical states, reducing search diversity and hindering objective optimization.
Although the AFM-stacked model shows regions of relatively small $R$ under this condition, the advantage of antiferromagnetic coupling in terms of solution quality appears limited, as discussed below.

Based on these observations, fair performance comparisons require $\mu \ge 5$, where all models stably produce feasible solutions.
Hereafter, we fix $\mu = 5$ unless otherwise noted.
The ability of the AFM-stacked model to achieve high $P_{\mathrm{Feasible}}$ at relatively small $\mu$ may be advantageous in hardware implementations subject to coefficient-range constraints or analog noise.~\cite{king2015performance, roch2023effect, mirkarimi2024quantum}

We emphasize that the choice of $\mu$ is not arbitrary.
The criterion $P_{\mathrm{Feasible}} \approx 1$ marks the regime where constraint violations are sufficiently suppressed, so that comparisons of solution quality based on the objective value become meaningful.
Below this threshold, infeasible configurations dominate the sampling, while above it the objective term governs the relative quality of feasible solutions.

\subsection{Dependence on the inter-replica coupling $J_{\mathrm{P}}$: stable operating region and solution quality}
\label{subsec:Jp_dependence}

The range of $J_{\mathrm{P}}$ is chosen relative to the typical coefficient scale of the problem Hamiltonian.
Values of $|J_{\mathrm{P}}|$ much smaller than the objective and constraint coefficients lead to weak inter-replica coupling, whereas excessively large $|J_{\mathrm{P}}|$ over-constrain replicas and suppress search diversity.
The explored range therefore spans the physically relevant regime from weak to strong coupling.

\begin{figure}[t]
    \centering
    \includegraphics[width=\linewidth]{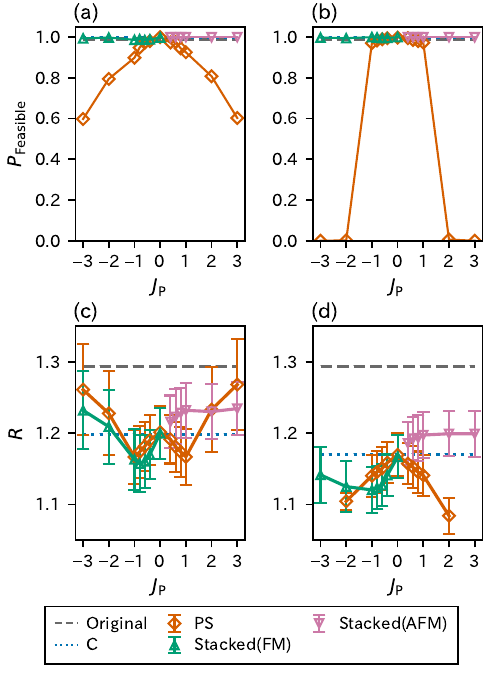}
    \caption{Dependence of $P_{\mathrm{Feasible}}$ and $R$ on $|J_{\mathrm{P}}|$ at $\mu=5$.
    (a)~$P_{\mathrm{Feasible}}$ for $P=10$.
    (b)~$P_{\mathrm{Feasible}}$ for $P=30$.
    (c)~$R$ for $P=10$.
    (d)~$R$ for $P=30$.
    Error bars indicate standard deviations.
    Lines connecting data points are guides to the eye.}
    \label{Fig:approximation_ratio_vs_Jp}
\end{figure}

We now examine the dependence on the inter-replica coupling $J_{\mathrm{P}}$, another key parameter governing model performance.
Fixing $\mu = 5$, we evaluate how $P_{\mathrm{Feasible}}$ and $R$ depend on $J_{\mathrm{P}}$ for $P=10$ and $P=30$.
Figure~\ref{Fig:approximation_ratio_vs_Jp} summarizes the results.
Because the C model and the original (single-replica) baseline contain no inter-replica couplings, their values are independent of $|J_{\mathrm{P}}|$ and appear as horizontal lines.
The PS model is insensitive to the sign of $J_{\mathrm{P}}$ and thus behaves symmetrically about $J_{\mathrm{P}}=0$.

Figures~\ref{Fig:approximation_ratio_vs_Jp}(a) and \ref{Fig:approximation_ratio_vs_Jp}(b) show that the stacked model maintains $P_{\mathrm{Feasible}} \approx 1.0$ throughout the range $-3 \le J_{\mathrm{P}} \le 3$.
Whether ferromagnetic coupling ($J_{\mathrm{P}}<0$) promotes synchronization or antiferromagnetic coupling ($J_{\mathrm{P}}>0$) introduces repulsion, the stacked topology does not impede constraint satisfaction.
This broad stable operating region constitutes a significant practical advantage for parameter tuning.
In contrast, for the PS model $P_{\mathrm{Feasible}}$ drops sharply with increasing $|J_{\mathrm{P}}|$; in particular, at $P=30$ we find $P_{\mathrm{Feasible}} \approx 0$ for $|J_{\mathrm{P}}| \ge 2$.
The PS model thus has a much narrower admissible $J_{\mathrm{P}}$ range than the stacked model, making parameter selection more challenging.

We next examine solution quality as measured by $R$.
Figure~\ref{Fig:approximation_ratio_vs_Jp}(c) shows that, for $P=10$, the stacked model attains its minimum $R$ near $J_{\mathrm{P}} \approx -1$.
In the antiferromagnetic regime ($J_{\mathrm{P}}>0$), $R$ increases and can exceed that of the C model.
Even in the ferromagnetic regime, excessively large $|J_{\mathrm{P}}|$ raises $R$, indicating that over-constraining replicas reduces search diversity and hinders optimization.
The PS model behaves similarly: $R$ is minimized around $|J_{\mathrm{P}}| \approx 1$ and increases for larger $|J_{\mathrm{P}}|$.

For $P=30$ [Fig.~\ref{Fig:approximation_ratio_vs_Jp}(d)], the PS model appears to achieve a low value of $R$ near $|J_{\mathrm{P}}| \approx 2$.
However, comparison with Fig.~\ref{Fig:approximation_ratio_vs_Jp}(b) reveals that $P_{\mathrm{Feasible}}$ is nearly zero in this region; the apparently low $R$ is therefore an artifact arising from a predominance of infeasible samples.
In contrast, in the ferromagnetic regime ($J_{\mathrm{P}}<0$) of the stacked model, $R$ decreases monotonically with increasing $|J_{\mathrm{P}}|$ (at least up to $|J_{\mathrm{P}}| \approx 2$) while maintaining $P_{\mathrm{Feasible}} \approx 1.0$.
This demonstrates that the stacked topology can convert cooperative inter-replica effects into improved solution quality without compromising constraint satisfaction.
In the antiferromagnetic regime ($J_{\mathrm{P}}>0$), constraints are still satisfied, but $R$ does not improve, suggesting that repulsion between adjacent replicas impedes convergence to the ground state.

In summary, the PS model is difficult to tune because of its narrow stable $J_{\mathrm{P}}$ range, whereas the stacked model achieves both constraint satisfaction and high solution quality over a broad $J_{\mathrm{P}}$ range.
Among the models considered, the ferromagnetically coupled stacked model exhibits the most practical overall performance.

\subsection{Scalability: dependence on the number of replicas $P$ and the number of MCMC steps $N_{\mathrm{Steps}}$}
\label{subsec:scalability}

\begin{figure}[t]
    \centering
    \includegraphics[width=\linewidth]{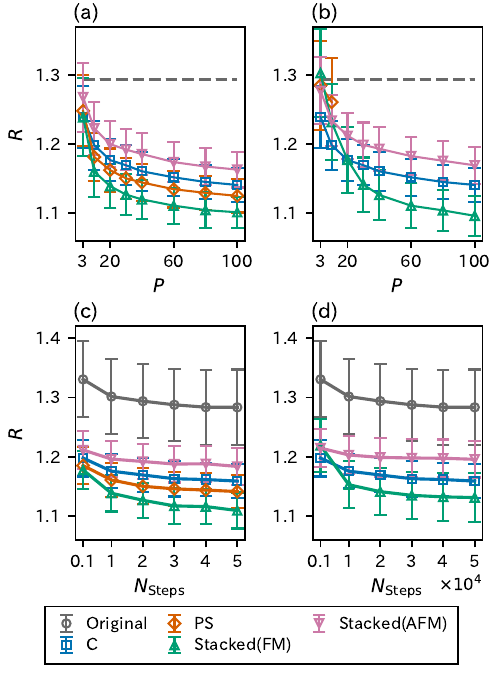}
    \caption{Dependence of $R$ on $P$ and $N_{\mathrm{Steps}}$ at $\mu=5$.
    (a)~$P$ dependence at $|J_{\mathrm{P}}|=0.6$.
    (b)~$P$ dependence at $|J_{\mathrm{P}}|=3$.
    (c)~$N_{\mathrm{Steps}}$ dependence at $|J_{\mathrm{P}}|=0.6$ with $P=30$.
    (d)~$N_{\mathrm{Steps}}$ dependence at $|J_{\mathrm{P}}|=3$ with $P=30$.
    Error bars indicate standard deviations.
    Lines connecting data points are guides to the eye.}
    \label{Fig:approximation_ratio_vs_P}
\end{figure}

We next evaluate scalability with respect to the number of replicas $P$ and the number of MCMC steps $N_{\mathrm{Steps}}$.
We fix $\mu=5$, at which all models stably yield feasible solutions.
For the inter-replica coupling, we consider two conditions: $|J_{\mathrm{P}}|=0.6$, where both replica-coupled models operate stably, and $|J_{\mathrm{P}}|=3$, which imposes stronger cooperative effects.
Figure~\ref{Fig:approximation_ratio_vs_P} shows the dependence of the approximation ratio $R$ on $P$ and $N_{\mathrm{Steps}}$.

In Fig.~\ref{Fig:approximation_ratio_vs_P}(a) ($|J_{\mathrm{P}}|=0.6$), $R$ decreases monotonically with increasing $P$ for all models, indicating that greater parallelism raises the probability of finding higher-quality solutions.
Among the models, the FM-stacked model yields the smallest $R$, i.e., the best solution quality.
In contrast, Fig.~\ref{Fig:approximation_ratio_vs_P}(b) ($|J_{\mathrm{P}}|=3$) shows that both the PS model and the FM-stacked model exhibit degraded $R$ at small $P$, suggesting that this coupling strength over-constrains the replicas and reduces search diversity.
Nevertheless, the performance of the FM-stacked model improves dramatically as $P$ increases, eventually reaching a solution quality comparable to that obtained at $|J_{\mathrm{P}}|=0.6$.
This indicates that the stacked topology can mitigate over-constraint through increased $P$, thereby extracting the benefit of ferromagnetic coupling.
By contrast, the PS model fails to yield feasible solutions for $P>10$ (hence no data points appear), implying that simply increasing $P$ does not prevent search breakdown.
The AFM-stacked model also shows decreasing $R$ with increasing $P$, but its performance remains consistently below that of the C model.

Next, we fix $P=30$ and examine convergence behavior as $N_{\mathrm{Steps}}$ is increased from $1000$ to $50000$.
Figures~\ref{Fig:approximation_ratio_vs_P}(c) and \ref{Fig:approximation_ratio_vs_P}(d) show the $N_{\mathrm{Steps}}$ dependence of $R$.
In Fig.~\ref{Fig:approximation_ratio_vs_P}(c) ($|J_{\mathrm{P}}|=0.6$), $R$ decreases with increasing $N_{\mathrm{Steps}}$ for all models, although the rate of improvement varies.
The FM-stacked model exhibits the steepest reduction in $R$, indicating that, given sufficient relaxation time, it converges efficiently to better solutions through cooperative interactions.
The PS model shows initial improvement but does not reach the final performance level of the FM-stacked model.
In Fig.~\ref{Fig:approximation_ratio_vs_P}(d) ($|J_{\mathrm{P}}|=3$), the FM-stacked model improves substantially as $N_{\mathrm{Steps}}$ increases, whereas the PS model fails to produce feasible solutions and effective search is not realized.

From the viewpoint of scalability with respect to both $P$ and $N_{\mathrm{Steps}}$, the FM-stacked model achieves the best overall balance between solution quality and stability.

\subsection{Size scaling: validation on multiple QAP instances}
\label{subsec:size_scaling}

In this subsection, we verify that the trends observed above are not specific to \texttt{tai12a} but persist for larger instances with substantially larger search spaces.
We compare performance on the QAPLIB instances \texttt{tai15a} ($L=15$), \texttt{tai17a} ($L=17$), and \texttt{tai20a} ($L=20$).

\begin{figure}[t]
    \centering
    \includegraphics[width=\linewidth]{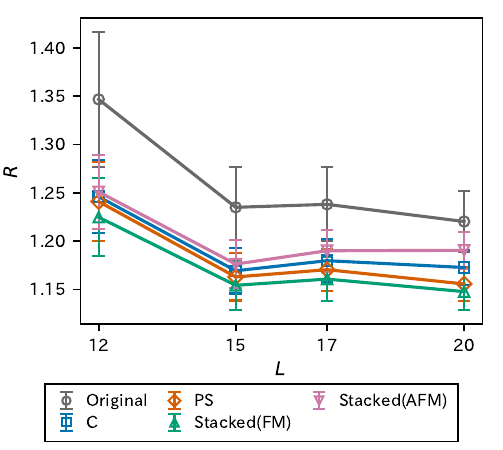}
    \caption{Dependence of $R$ on the QAP size $L$.
    Parameters are $P=10$, $\mu=8$, and $|J_{\mathrm{P}}|=1$.
    Error bars indicate standard deviations.
    Lines connecting data points are guides to the eye.}
    \label{Fig:approximation_ratio_vs_L}
\end{figure}

Figure~\ref{Fig:approximation_ratio_vs_L} shows the approximation ratio $R$ as a function of $L$.
We fix $P=10$ and set $\mu=8$ and $|J_{\mathrm{P}}|=1$, where $\mu=8$ is chosen as a common value at which all models stably obtain feasible solutions for all problem sizes considered.
The FM-stacked model consistently achieves the smallest $R$ across all problem sizes, and the performance gap relative to the C, PS, and AFM-stacked models widens as $L$ increases.
This suggests that cooperative search in the FM-stacked model becomes relatively more effective as the problem difficulty grows.

\begin{figure}[t]
    \centering
    \includegraphics[width=\linewidth]{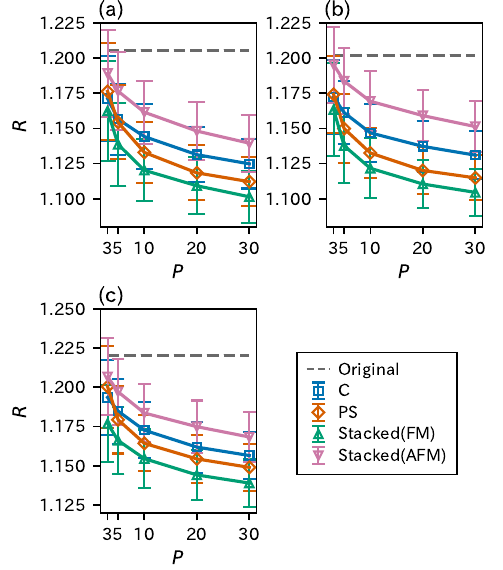}
    \caption{Dependence of $R$ on $P$ for QAP instances with $L=15$, $17$, and $20$ at $|J_{\mathrm{P}}|=0.6$.
    (a)~$L=15$ with $\mu=6$.
    (b)~$L=17$ with $\mu=6$.
    (c)~$L=20$ with $\mu=8$.
    Error bars indicate standard deviations.
    Lines connecting data points are guides to the eye.}
    \label{Fig:scalability_large_L}
\end{figure}

We further examine scalability with increasing $P$ on these larger instances.
Figure~\ref{Fig:scalability_large_L} shows the $P$ dependence of $R$ for $L=15$, $17$, and $20$.
The inter-replica coupling is fixed at $|J_{\mathrm{P}}|=0.6$ for all sizes to test whether the stacked model operates robustly with a common $J_{\mathrm{P}}$ independent of problem size.
The penalty coefficient $\mu$ is set to the minimum value yielding $P_{\mathrm{Feasible}}=1.0$: $\mu=6$ for $L=15$ and $17$, and $\mu=8$ for $L=20$.
Common trends are observed across all problem sizes.
In particular, the FM-stacked model reduces $R$ most steeply with increasing $P$ at every size, demonstrating high scalability even as the problem grows.
These results further confirm that, for QAP, the FM-stacked model is the most practical choice: it maintains constraint satisfaction over a broad parameter region while effectively converting increased parallelism into improved solution quality.

We note that the largest instances examined here ($L \le 20$) are limited by the computational cost of systematic parameter scans rather than by conceptual constraints of the replica-coupled models.
Our focus is therefore on elucidating qualitative and structural scaling trends, rather than on pushing the maximum solvable size.
Because the stacked and PS models introduce only $\mathcal{O}(P)$ additional couplings per variable, their relative computational overhead scales linearly with problem size, suggesting that the structural differences identified here are expected to persist at larger $L$.

\section{Mechanism: Replica Correlations and Bit Configurations}
\label{Sec:mechanism}

In the previous section, we compared optimization performance and robustness primarily through the feasibility rate $P_{\mathrm{Feasible}}$ and the approximation ratio $R$.
We found that the stacked model (especially with ferromagnetic coupling) stably satisfies constraints and improves solution quality over a wide parameter range, whereas the PS model is vulnerable to increases in the number of replicas $P$, exhibiting performance saturation or breakdown.
In this section, we analyze SA samples from the viewpoints of inter-replica correlations and bit statistics to elucidate the physical origin of these contrasting behaviors, and discuss how model topology affects the balance between cooperation and diversity.

\subsection{Qualitative assessment of inter-replica cooperation: 1--1 correlation}
\label{subsec:1-1correlation}

A naive approach to quantifying inter-replica cooperation is to measure the agreement of corresponding bits or spin correlations.
However, for problems with extremely sparse solutions such as QAP, where the vast majority of bits are $0$, simple correlations tend to be overestimated.
Even when replicas do not share essential solution information (i.e., which positions take the value $1$), agreement on zeros dominates and yields spuriously high correlation values.
Simple correlations therefore cannot distinguish genuine cooperation (shared solution structure) from trivial agreement on zeros.

In such sparse constrained problems, the physically relevant notion of cooperation is not overall similarity, but whether replicas consistently identify and share the locations of the few active variables.
Measures such as Hamming distance or spin--spin correlation are dominated by zero-valued bits and therefore do not provide a clear indicator of coordinated search behavior in this setting.

To extract only meaningful cooperation, we define a conditional agreement rate restricted to bits for which at least one replica takes the value $1$, thereby excluding trivial agreement on zeros.
Let $N^{(p,p+1)}_{ab}$ ($a,b\in\{0,1\}$) denote the number of bit pairs $(x^{(p)}_{i,k}, x^{(p+1)}_{i,k})$ in state $(a,b)$ between adjacent replicas $p$ and $p+1$.
The 1--1 correlation $\langle S\rangle_{1}$ is defined as
\begin{align}
    \langle S\rangle_{1}
    =
    \frac{1}{P}\sum_{p=1}^{P}
    \frac{N^{(p,p+1)}_{11}}
    {N^{(p,p+1)}_{11}+N^{(p,p+1)}_{10}+N^{(p,p+1)}_{01}},
    \label{eq:S1_def}
\end{align}
where the replica index is taken modulo $P$ (i.e., $P+1 \to 1$).
This quantity is equivalent to the Jaccard index and measures how strongly replicas share essential solution information, namely, which positions are set to $1$.
By isolating the overlap of active variables, $\langle S\rangle_{1}$ captures the aspect of cooperation that is directly relevant to coordinated exploration of sparse feasible solutions.
A larger $\langle S\rangle_{1}$ indicates stronger sharing of solution structure between adjacent replicas, whereas a smaller value indicates more independent or effectively repulsive exploration.

\subsubsection{Dependence on $\mu$: relationship between constraint dominance and cooperation}

\begin{figure}[t]
    \centering
    \includegraphics[width=\linewidth]{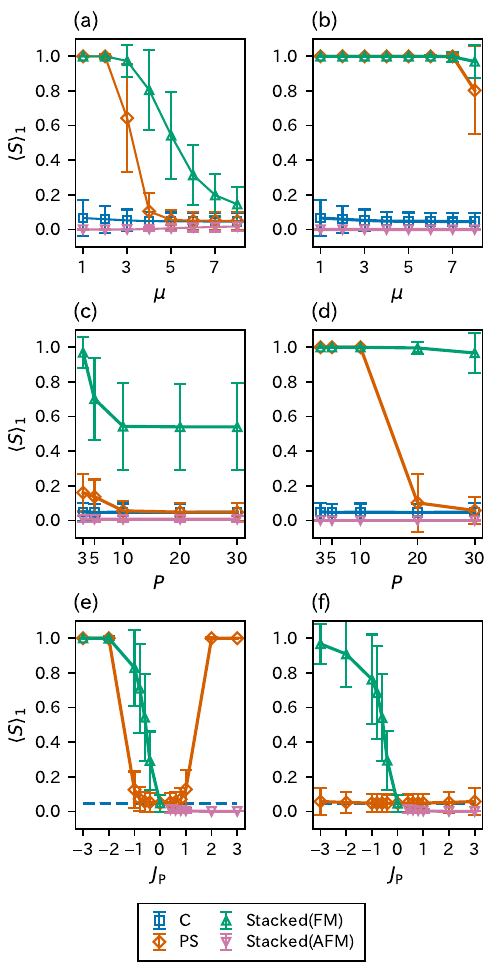}
    \caption{The 1--1 correlation $\langle S\rangle_{1}$.
    (a)~$\mu$ dependence at $P=10$ and $|J_{\mathrm{P}}|=0.6$.
    (b)~$\mu$ dependence at $P=10$ and $|J_{\mathrm{P}}|=3$.
    (c)~$P$ dependence at $\mu=5$ and $|J_{\mathrm{P}}|=0.6$.
    (d)~$P$ dependence at $\mu=5$ and $|J_{\mathrm{P}}|=3$.
    (e)~$J_{\mathrm{P}}$ dependence at $\mu=5$ and $P=10$.
    (f)~$J_{\mathrm{P}}$ dependence at $\mu=5$ and $P=30$.
    Error bars indicate standard deviations.
    Lines connecting data points are guides to the eye.}
    \label{Fig:jaccard}
\end{figure}

We first examine the behavior of $\langle S\rangle_{1}$ as $\mu$ varies, to clarify the relationship between energetic dominance of the constraint term and the quality of inter-replica cooperation.
We fix $P=10$ and compare two coupling strengths, $|J_{\mathrm{P}}|=0.6$ and $|J_{\mathrm{P}}|=3$.
Figures~\ref{Fig:jaccard}(a) and \ref{Fig:jaccard}(b) show the $\mu$ dependence of $\langle S\rangle_{1}$.

In Fig.~\ref{Fig:jaccard}(a) ($|J_{\mathrm{P}}|=0.6$), the FM-stacked and PS models exhibit large values $\langle S\rangle_{1} = 0.8$--$1.0$ in the low-$\mu$ regime ($\mu < 3$).
Here the constraint term is weak and the $J_{\mathrm{P}}$ interaction becomes relatively dominant. 
Each replica lowers its energy by aligning with neighbors by placing $1$ at the same positions rather than satisfying constraints.
Consequently, strong inter-replica alignment emerges even though constraints are not necessarily satisfied.

As $\mu$ increases, $\langle S\rangle_{1}$ gradually decreases.
This reflects the growing dominance of the constraint term: replicas begin to prioritize individual constraint satisfaction over alignment, increasing diversity across replicas.

In Fig.~\ref{Fig:jaccard}(b) ($|J_{\mathrm{P}}|=3$), the FM-stacked and PS models maintain $\langle S\rangle_{1} \approx 1.0$ even at large $\mu$, suggesting that strong coupling excessively constrains the system and freezes nearly all layers into identical states.
In contrast, the AFM-stacked and C models remain at $\langle S\rangle_{1} \approx 0$ throughout.
For the C model this is expected because replicas are independent.
For the AFM-stacked model, the coupling disfavors matching the positions of $1$ between adjacent replicas, preventing formation of a common solution structure.

\subsubsection{Dependence on $P$: collapse of cooperation in PS versus persistence in stacked}

We now fix $\mu=5$ (a stable operating regime) and examine how $\langle S\rangle_{1}$ behaves as $P$ varies.
Figures~\ref{Fig:jaccard}(c) and \ref{Fig:jaccard}(d) show the $P$ dependence of $\langle S\rangle_{1}$.

For the PS model, $\langle S\rangle_{1}$ decreases monotonically with increasing $P$ and converges to nearly zero at both $|J_{\mathrm{P}}|=0.6$ and $|J_{\mathrm{P}}|=3$.
This indicates that increasing parallelism in the PS topology prevents replicas from maintaining effective cooperation, ultimately leaving them unable to share essential solution information.

The FM-stacked model exhibits a contrasting trend.
In Fig.~\ref{Fig:jaccard}(c) ($|J_{\mathrm{P}}|=0.6$), $\langle S\rangle_{1}$ decreases with $P$ but does not vanish; it saturates at a finite value (${\sim}0.4$--$0.5$).
This can be attributed to the locality of the stacked topology: each replica interacts only with its neighbors, so the character of the interaction does not change abruptly as $P$ increases, allowing adjacent replicas to retain a tendency to share solution structure.
In Fig.~\ref{Fig:jaccard}(d) ($|J_{\mathrm{P}}|=3$), $\langle S\rangle_{1}$ remains above $0.9$ almost independently of $P$, indicating that the replica ensemble is strongly unified by the strong coupling.
When such cooperation is preserved, increasing $P$ effectively increases the number of search trials, consistent with the improvement in solution quality observed in the previous section.
The C and AFM-stacked models remain at $\langle S\rangle_{1} \approx 0$ throughout, confirming that no cooperation in terms of aligning the positions of $1$ is present.

\subsubsection{Dependence on $J_{\mathrm{P}}$: topology-dependent divergence at large $P$}

Finally, we fix $\mu=5$ and examine the behavior of $\langle S\rangle_{1}$ as $J_{\mathrm{P}}$ varies.
Figures~\ref{Fig:jaccard}(e) and \ref{Fig:jaccard}(f) show results for $P=10$ and $P=30$, respectively.

At $P=10$ [Fig.~\ref{Fig:jaccard}(e)], both the PS and FM-stacked models show increasing $\langle S\rangle_{1}$ with $|J_{\mathrm{P}}|$ in the ferromagnetic regime ($J_{\mathrm{P}}<0$), indicating strengthened cooperation.
Thus, at moderate parallelism ($P \sim 10$), both topologies can utilize $J_{\mathrm{P}}$ to enhance cooperation.

At $P=30$ [Fig.~\ref{Fig:jaccard}(f)], the difference becomes decisive.
The FM-stacked model behaves similarly to the $P=10$ case: $\langle S\rangle_{1}$ increases with $|J_{\mathrm{P}}|$ in the ferromagnetic regime, maintaining inter-replica correlation.
In contrast, the PS model remains essentially pinned at $\langle S\rangle_{1} \approx 0$ across the entire $J_{\mathrm{P}}$ range, and increasing $|J_{\mathrm{P}}|$ does not restore cooperation.
This indicates that beyond a certain level of parallelism, the PS topology cannot establish effective cooperation by tuning $J_{\mathrm{P}}$ alone, revealing an intrinsic limitation imposed by the centralized coupling topology.

\subsection{Mechanism of cooperation collapse in the PS model}
\label{subsec:PS_collapse}

The preceding analysis showed that, in the PS model, the 1--1 correlation $\langle S\rangle_{1}$ vanishes as $P$ increases: replicas fail to share solution structure.
In this subsection, we identify the underlying cause of this collapse by focusing on the PS layer and analyzing bit configurations and mean bit values.

\subsubsection{Bit configurations: role of the PS layer at different $\mu$}

\begin{figure*}[t]
    \centering
    \includegraphics[width=\linewidth]{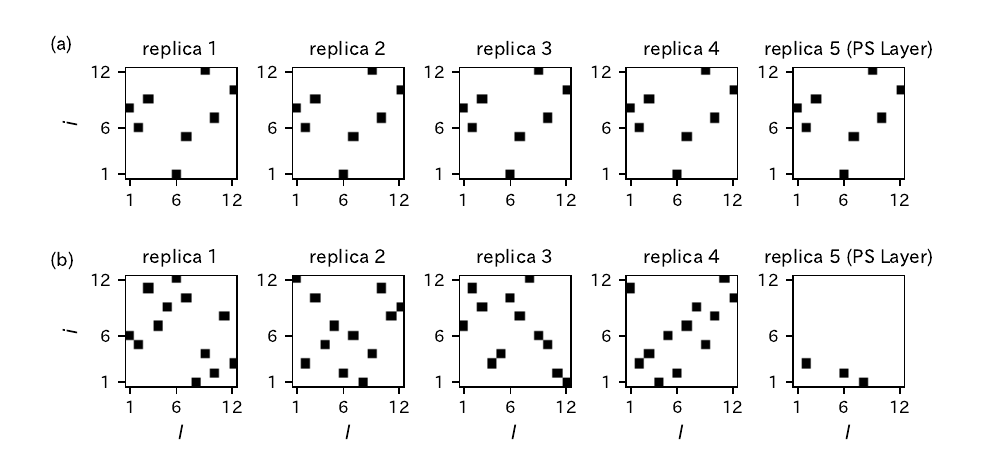}
    \caption{Bit configurations obtained by SA for QAP (\texttt{tai12a}, $L=12$) with $|J_{\mathrm{P}}|=0.6$ and $P=5$.
    (a)~$\mu=1$.
    (b)~$\mu=8$.
    White and black indicate bits $0$ and $1$, respectively.
    Replica $p=5$ is the PS layer.}
    \label{Fig:spin_configuration}
\end{figure*}

To visualize the structural characteristics of the PS model, we fix $|J_{\mathrm{P}}|=0.6$ and $P=5$, and inspect final bit configurations obtained by SA at different $\mu$.
Figure~\ref{Fig:spin_configuration} shows representative samples for (a)~$\mu=1$ and (b)~$\mu=8$; replica $p=5$ corresponds to the PS layer.

At $\mu=1$, the constraint term is weak and the $J_{\mathrm{P}}$ interaction dominates.
All replicas, including the PS layer, tend to align to an identical configuration.
However, because constraints are not sufficiently enforced, strong alignment can occur without guaranteeing feasibility.

At $\mu=8$, the constraint term sets a larger energy scale.
Each replica then prioritizes constraint satisfaction and explores more independently, increasing diversity.
The PS layer effectively aggregates these diverse (typically feasible) solutions through averaging; owing to the sparsity of QAP (most bits are $0$), the PS layer becomes dominated by zeros.
Consequently, as $\mu$ increases, the PS model effectively approaches the behavior of independent replicas (C model).

\subsubsection{Dependence of the PS-layer mean bit value on $\mu$}

\begin{figure}[t]
    \centering
    \includegraphics[width=\linewidth]{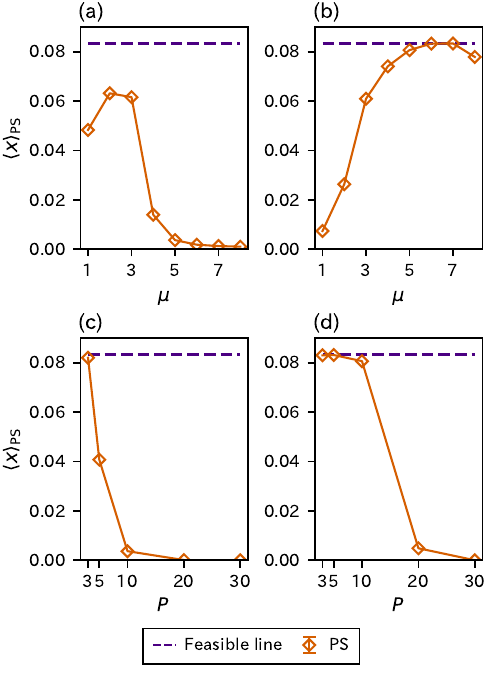}
    \caption{Mean bit value in the PS layer, $\langle x\rangle_{\mathrm{PS}}$.
    (a)~$\mu$ dependence at $P=10$ and $|J_{\mathrm{P}}|=0.6$.
    (b)~$\mu$ dependence at $P=10$ and $|J_{\mathrm{P}}|=3$.
    (c)~$P$ dependence at $\mu=5$ and $|J_{\mathrm{P}}|=0.6$.
    (d)~$P$ dependence at $\mu=5$ and $|J_{\mathrm{P}}|=3$.
    Dashed lines indicate the theoretical value for feasible solutions.
    Error bars indicate standard deviations.
    Lines connecting data points are guides to the eye.}
    \label{Fig:magnetization_PS}
\end{figure}

We next examine the $\mu$ dependence of the mean bit value in the PS layer, $\langle x\rangle_{\mathrm{PS}}$.
Figures~\ref{Fig:magnetization_PS}(a) and \ref{Fig:magnetization_PS}(b) show $\langle x\rangle_{\mathrm{PS}}$ versus $\mu$ for $P=10$.
The dashed line indicates the theoretical value for feasible solutions: with $L=12$ and $N=L^{2}=144$, exactly $L$ bits equal $1$, giving $\langle x\rangle = L/N \approx 0.083$.

In Fig.~\ref{Fig:magnetization_PS}(a) ($|J_{\mathrm{P}}|=0.6$), $\langle x\rangle_{\mathrm{PS}}$ decreases with increasing $\mu$ and approaches zero.
As $\mu$ grows, each replica increasingly prioritizes constraint satisfaction and tends to settle into distinct feasible solutions.
Because the PS layer aggregates (averages over) these diverse configurations, it is statistically pulled toward zero.
The high $P_{\mathrm{Feasible}}$ observed at large $\mu$ is thus achieved not through enforced alignment by the PS layer, but as an accumulation of nearly independent searches.

In Fig.~\ref{Fig:magnetization_PS}(b) ($|J_{\mathrm{P}}|=3$), increasing $\mu$ drives $\langle x\rangle_{\mathrm{PS}}$ toward the theoretical feasible value.
Here the energy scale of $J_{\mathrm{P}}$ is large, and all layers tend to align to a single feasible solution, consistent with improved $P_{\mathrm{Feasible}}$.

\subsubsection{Dependence of the PS-layer mean bit value on $P$: information loss through averaging}

Finally, we fix $\mu=5$ and analyze how $\langle x\rangle_{\mathrm{PS}}$ changes as $P$ increases.
Figures~\ref{Fig:magnetization_PS}(c) and \ref{Fig:magnetization_PS}(d) show results for $|J_{\mathrm{P}}|=0.6$ and $|J_{\mathrm{P}}|=3$, respectively.
In both cases, $\langle x\rangle_{\mathrm{PS}}$ decreases toward zero as $P$ increases.
At $|J_{\mathrm{P}}|=0.6$, deviation from the feasible-line value begins at relatively small $P$.
At $|J_{\mathrm{P}}|=3$, $\langle x\rangle_{\mathrm{PS}}$ remains near the feasible value for $P \lesssim 10$ but eventually decreases as $P$ grows further.

This behavior can be understood from the sparsity of QAP solutions combined with many-replica averaging in the PS layer.
The PS layer receives the average tendency of corresponding bits across all replicas.
Because feasible solutions are sparse, unless the positions of $1$ coincide across replicas, averaging over many replicas drives the result toward zero.
At small $P$, accidental overlap of $1$ bits or strong coupling can produce alignment.
As $P$ increases, however, the information carried by the minority of $1$ bits is diluted by the majority of zeros, pulling the PS layer toward an all-zero configuration.

Once the PS layer becomes biased toward zero, each replica experiences two competing pressures: (i)~to place $1$ bits for constraint satisfaction, and (ii)~suppression toward zero mediated by the PS layer.
As $P$ grows, pressure~(ii) strengthens statistically, and the PS layer ceases to function as an effective guide for sharing solution structure across replicas.
Replicas may place $1$ bits sporadically but fail to coordinate their positions, leading to the disappearance of cooperation ($\langle S\rangle_{1} \to 0$).
Under some conditions, this can even hinder feasibility, yielding a regime where replicas are disordered and fail to satisfy constraints.

We therefore conclude that the performance degradation of the PS model at large $P$ arises not from excessive alignment, but from the loss of sparse solution information through many-replica averaging.
This averaging deprives the PS topology of an effective mechanism for establishing inter-replica cooperation.

We emphasize that the mechanism identified here is not specific to QAP.
Rather, it is a generic consequence of replica averaging under constraints that induce sparse feasible solutions.
For optimization problems with exactly-one or one-hot type constraints, feasible configurations typically contain only a small fraction of active variables.
When different replicas settle into distinct feasible solutions, averaging their configurations in a centralized auxiliary layer inevitably drives the mean field toward zero, thereby washing out essential solution information.
This argument applies broadly to constrained optimization problems with sparse feasible structures, such as assignment, matching, and scheduling problems.

\subsection{Constraint-satisfaction mechanism of the AFM-stacked model at low $\mu$}
\label{subsec:AFM_mechanism}

Finally, we interpret why the AFM-stacked model exhibits higher $P_{\mathrm{Feasible}}$ than the other models in the low-$\mu$ regime, based on the mean bit value averaged over all replicas.

In QAP, a typical constraint-violating local minimum is the all-zero state, in which all bits equal $0$.
Such a state arises easily in highly sparse problems.
The sign of the inter-replica coupling is expected to strongly affect the stability of this state.

\begin{figure}[t]
    \centering
    \includegraphics[width=\linewidth]{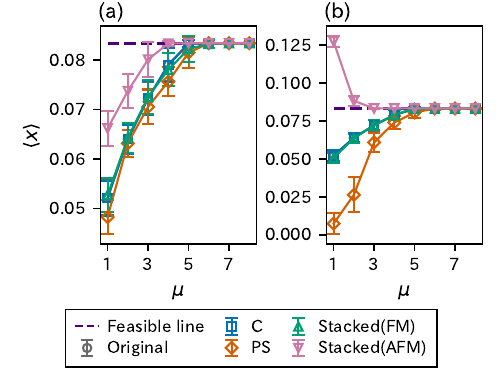}
    \caption{Dependence of the mean bit value $\langle x\rangle$ over all replicas on $\mu$ for $P=10$.
    (a)~$|J_{\mathrm{P}}|=0.6$.
    (b)~$|J_{\mathrm{P}}|=3$.
    Dashed lines indicate the theoretical value for feasible solutions.
    Error bars indicate standard deviations.
    Lines connecting data points are guides to the eye.}
    \label{Fig:magnetization_vs_mu}
\end{figure}

Figure~\ref{Fig:magnetization_vs_mu} shows the $\mu$ dependence of the mean bit value $\langle x\rangle$ over all replicas for $P=10$.
In Fig.~\ref{Fig:magnetization_vs_mu}(a) ($|J_{\mathrm{P}}|=0.6$), within the low-$\mu$ regime ($\mu < 4$), the FM-stacked, PS, and C models all yield $\langle x\rangle$ far below the feasible-solution line, approaching zero.
This indicates that the energetic drive toward constraint satisfaction is too weak, allowing the system to become trapped in the all-zero state.
In contrast, the AFM-stacked model shows a suppressed reduction of $\langle x\rangle$ in the same regime, remaining closer to the feasible-line value.
This can be attributed to the antiferromagnetic interaction, which makes alignment of zeros between adjacent replicas energetically unfavorable, thereby destabilizing the all-zero state.

The difference becomes more pronounced in Fig.~\ref{Fig:magnetization_vs_mu}(b) ($|J_{\mathrm{P}}|=3$).
While the other models exhibit $\langle x\rangle$ below the feasible line in the low-$\mu$ regime, the AFM-stacked model clearly exceeds the feasible-line value.
This implies that the number of $1$ bits is larger than in a feasible solution, suggesting that strong antiferromagnetic coupling locally generates $1$ bits to avoid zero alignment.
With this repulsive assistance from antiferromagnetic coupling, feasible solutions can be obtained with relatively high probability even at $\mu$ values where other models fail to achieve feasibility.

In summary, the AFM-stacked model improves feasibility by avoiding the all-zero state; however, this does not guarantee convergence to the true optimum.
This interpretation is consistent with the observation that, despite its high $P_{\mathrm{Feasible}}$, the AFM-stacked model does not readily translate into improvements in the approximation ratio $R$.

\section{Practical Guidelines}
\label{Sec:practicalguidelines}

Based on the findings of this study, we summarize practical guidelines for selecting a replica-coupled model and tuning its parameters when applying Ising machines (or equivalent annealing-type solvers) to real-world problems.
Because our analysis isolates structural effects using noise-free SA, the guidelines below should be viewed as a baseline derived from the manner in which the inter-replica coupling topology shapes search behavior, rather than as a direct quantification of noise-suppression benefits.

\subsection{Model selection: applicability of the PS and stacked models}
\label{subsec:model_selection}

We recommend the FM-stacked model when targeting (i)~large-scale parallelism (large $P$) and/or (ii)~constrained optimization problems with sparse feasible structures.
As demonstrated in this study, the FM-stacked model improves solution quality while maintaining constraint satisfaction over a wide $(\mu, J_{\mathrm{P}})$ range and converts increases in $P$ and $N_{\mathrm{Steps}}$ into performance gains in a largely monotonic manner.
This broad stable operating region and favorable scalability reduce both tuning cost and operational risk.

The PS model can yield performance improvements through cooperative effects when the number of replicas is small.
However, cooperation tends to collapse as $P$ increases, and there emerges a regime in which effective cooperation cannot be established regardless of how $J_{\mathrm{P}}$ is tuned.
For constrained problems with sparse solutions, many-replica averaging causes the PS layer to lose its guiding function, making the PS model unsuitable for applications that require large $P$.
If the PS model is adopted, it is advisable to keep $P$ small and to verify beforehand whether a parameter region exists in which both $P_{\mathrm{Feasible}}$ and solution quality are satisfactory.

The AFM-stacked model has the advantage of achieving feasible solutions even at low $\mu$.
For the QAP studied here, however, this advantage does not readily translate into improved solution quality.
Hence, the AFM-stacked model is a reasonable choice when the primary goal is to secure feasibility, whereas the FM-stacked model should be preferred when solution quality is the primary concern.

\subsection{Recommended parameter-tuning procedure}
\label{subsec:parameter_tuning}

We provide a practical workflow for parameter tuning when adopting the stacked model, particularly the FM-stacked variant.
The primary tunable parameters are the constraint-penalty coefficient $\mu$, the inter-replica coupling $J_{\mathrm{P}}$, the number of replicas $P$, and (if resources permit) the number of MCMC steps $N_{\mathrm{Steps}}$.
The procedure is designed to first secure constraint satisfaction and then improve solution quality in a controlled manner.

\paragraph{Step 1: Determine $\mu$.}
Using either the original (single-replica) setting or the C model, estimate the minimum $\mu$ required for reliable constraint satisfaction on the target problem.
Identify the smallest $\mu$ at which $P_{\mathrm{Feasible}}$ becomes sufficiently high (e.g., $P_{\mathrm{Feasible}} \approx 1$), then adopt a value with an additional margin.
Although the AFM-stacked model can achieve feasibility at smaller $\mu$, it is generally safer to assume the FM-stacked model and choose $\mu$ with adequate headroom when high-quality solutions are the priority.

\paragraph{Step 2: Set $J_{\mathrm{P}}$.}
Avoid choosing $|J_{\mathrm{P}}|$ excessively large relative to the coefficient scale of the problem Hamiltonian; start from a modest value.
Increase $|J_{\mathrm{P}}|$ gradually within the range that preserves high $P_{\mathrm{Feasible}}$, and search for a regime where the approximation ratio improves.
Our results indicate that the stacked model maintains feasibility over a broad $J_{\mathrm{P}}$ range and is likely to operate stably without fine-grained tuning.
However, excessively large $|J_{\mathrm{P}}|$ can reduce search diversity at small $P$, degrading performance.
The strategy that ``stronger coupling is always better'' should therefore be avoided.

\paragraph{Step 3: Choose $P$.}
Within available runtime and hardware constraints, set $P$ as large as possible.
For the FM-stacked model, increasing $P$ does not readily induce cooperation collapse and is effectively converted into improved solution quality.
Thus, $P$ serves as a primary lever for improving performance.

\paragraph{Step 4: Choose $N_{\mathrm{Steps}}$.}
If feasible, increase $N_{\mathrm{Steps}}$ to provide sufficient relaxation time.
The FM-stacked model tends to improve continuously with increasing $N_{\mathrm{Steps}}$, making this an effective option when additional computational resources are available.

\paragraph{Implications for hardware implementations.}
Although the present study employs simulated annealing as a hardware-noise-free testbed, the structural insights obtained here are directly relevant to the design and operation of physical Ising machines.
In analog quantum annealers and coherent Ising machines, increasing the number of replicas primarily translates into an increase in the number of physical spins, while the annealing time per run is not expected to scale proportionally because all spins evolve in parallel.
From this perspective, replica-coupled models with decentralized interaction topologies, such as the stacked model, are particularly attractive for large-scale parallelization.

In practical hardware implementations, embedding replica-coupled models onto restricted connectivity graphs (e.g., Chimera or Pegasus architectures) introduces additional constraints.
Centralized coupling schemes, such as the PS model, require long-range interactions that can exacerbate embedding overhead, whereas the local connectivity of the stacked model is more naturally compatible with hardware-limited graphs.

\subsection{Operational considerations}
\label{subsec:operational}

We conclude with several operational remarks:
\begin{itemize}
    \item For constrained problems, high $P_{\mathrm{Feasible}}$ is a prerequisite; comparisons of approximation ratio or objective value should be made only when feasibility is ensured.
    \item If the PS model is adopted, large-$P$ operation requires caution because cooperation collapse (information loss through PS-layer averaging) can occur. Workflows that assume scaling with $P$ need careful validation.
    \item The AFM-stacked model may be suitable when the primary goal is to secure feasibility at low $\mu$, but it does not guarantee improved solution quality. The choice between FM and AFM coupling should be made according to the application objective.
\end{itemize}

Following these guidelines should help practitioners reliably extract the benefits of replica-coupled models in terms of both constraint satisfaction and solution quality.
\section{Conclusion}
\label{Sec:conclusion}

In this study, we reframed error-mitigation methods for Ising machines as a structural design problem of replica-coupled Ising models, rather than merely as noise-suppression techniques.
We systematically compared the penalty-spin (PS) model and the stacked model, analyzing how their structural differences affect optimization performance and search robustness.
To isolate the intrinsic effect of model topology on search dynamics, we employed simulated annealing (SA) as an idealized, noise-free evaluation platform.
As a benchmark, we focused on the quadratic assignment problem (QAP), which features strong assignment constraints and sparse solution structure.

Our numerical results show that the stacked model, particularly the ferromagnetically coupled variant (FM-stacked), stably maintains constraint satisfaction while improving solution quality over a broad range of the penalty coefficient $\mu$ and the inter-replica coupling $J_{\mathrm{P}}$.
The FM-stacked model converts increases in the number of replicas $P$ and the number of annealing steps $N_{\mathrm{Steps}}$ into consistent gains in solution quality, exhibiting favorable scalability as problem size grows.
In contrast, the PS model can exhibit cooperative benefits at small $P$, but rapidly loses inter-replica cooperation as $P$ increases, leading to unstable behavior in both feasibility and solution quality.

Microscopic analyses based on inter-replica correlations and bit-level statistics clarified the origin of these contrasting behaviors.
In the PS model, information from many replicas is averaged in the penalty-spin layer; for sparse constrained problems such as QAP, this averaging washes out essential solution information, and the PS layer ceases to function as an effective guide for coordination.
Consequently, under large-parallelism conditions, replicas fail to coordinate and feasibility itself becomes difficult to maintain.
In the stacked model, by contrast, local interactions between adjacent replicas preserve partial sharing of solution structure even as $P$ increases, sustaining a balance between cooperation and diversity.
This topological difference is the decisive factor underlying the gap in search robustness between the two models.

We also found that the AFM-stacked model tends to avoid the all-zero state in the low-$\mu$ regime, enabling it to obtain feasible solutions with relatively high probability.
However, the repulsive nature of antiferromagnetic coupling does not necessarily promote convergence toward the true optimum, and the AFM-stacked model does not match the FM-stacked model in improving solution quality.
Thus, while antiferromagnetic coupling can aid feasibility, caution is required if it is employed as the primary mechanism for high-quality optimization.

In summary, our results demonstrate that the effectiveness of error-mitigation approaches depends not only on potential noise-suppression effects but also strongly on the topology of inter-replica couplings.
By contrasting the centralized coupling architecture of the PS model with the decentralized topology of the stacked model, this work provides design guidelines for large-scale parallelization and constrained optimization.

Several directions remain for future work.
First, it is important to investigate how the structural characteristics identified here manifest under realistic hardware conditions that include control errors and thermal noise.
Second, extending the analysis to broader classes of constrained optimization problems would be valuable.
Third, examining connections to quantum annealing and quantum Monte Carlo dynamics is of interest.
The perspective developed here, treating error mitigation as structural design rather than solely as noise countermeasures, is expected to provide useful guidance for the design and operation of future Ising machines and annealing-based computing systems.

\section*{Acknowledgments}
This work was partially supported by the Japan Society for the Promotion of Science (JSPS) KAKENHI (Grant Number JP23H05447), the Council for Science, Technology, and Innovation (CSTI) through the Cross-ministerial Strategic Innovation Promotion Program (SIP), ``Promoting the application of advanced quantum technology platforms to social issues'' (Funding agency: QST), Japan Science and Technology Agency (JST) (Grant Number JPMJPF2221). The computations in this work were partially performed using the facilities of the Supercomputer Center, the Institute for Solid State Physics, The University of Tokyo. S. Tanaka wishes to express gratitude to the World Premier International Research Center Initiative (WPI), MEXT, Japan, for supporting the Human Biology-Microbiome-Quantum Research Center (Bio2Q).

\section*{Author contributions}
T.~Abe and K.~Hino contributed equally to this work.

\bibliographystyle{jpsj}
\bibliography{references}

\end{document}